\documentclass[12pt]{article}

\usepackage{amsmath,amssymb}
\usepackage{graphicx}
\usepackage{wrapfig}
\usepackage{multirow}
\usepackage{float}
\usepackage[usenames]{color}
\usepackage{indentfirst}  %package to shift the first paragraph
\usepackage{cite}
\usepackage{hyperref}
\hypersetup{
    colorlinks=true, %set true if you want colored links
    linkcolor=blue,  %choose some color if you want links to stand out
    citecolor=blue,
}
\allowdisplaybreaks[4]

\makeatletter

\@addtoreset{equation}{section}
\makeatother

\newcommand{{\Slashp}}{p\!\!\!\!\!\big/}
\newcommand{{\Slashq}}{q\!\!\!\!\!\big/}

\makeatletter
\newcommand{\figcaption}[1]{\def\@captype{figure}\caption{#1}}
\newcommand{\tblcaption}[1]{\def\@captype{table}\caption{#1}}
\makeatother
\begin{document}

\title{Numerical Analysis of Coleman-de Luccia Tunneling}

\author{%       %Use \scshape  for the family name
Yuhei \textsc{Goto}\footnote{E-mail: 14st302a@shinshu-u.ac.jp} ~~and~ 
%Masaatsu \textsc{Horikoshi}\footnote{E-mail: 14st307b@azusa.shinshu-u.ac.jp}~~,~
Kazumi \textsc{Okuyama}\footnote{E-mail: kazumi@azusa.shinshu-u.ac.jp}\\
{\it Department of Physics, Shinshu University, }\\
{\it Matsumoto 390-8621, Japan}
}

%\date{%      %Editorial Office will fill in this.
%May 14, 2015}

\maketitle
\begin{abstract}
%We know that vacuum decay occurs when potential has a true and false vacuum. 
%And, vacuum decay including the effect of gravity is important 
%for the study of the early universe and the cosmic landscape. 
We study the false vacuum decay of a single scalar field $\phi$
coupled to gravity 
described by the Coleman-de Luccia (CdL) instanton. 
We show that it is possible to numerically calculate the bounce factor, 
which is related to the CdL tunneling rate, 
without using the thin-wall approximation. 
In this paper, we consider $1/\cosh(\phi)$- and $\cos(\phi)$-type potential as examples, 
which have cosmological and phenomenological applications. 
Especially, in the $\cos(\phi)$-type potential 
we show that the range of values in which axion decay constant can take is restricted
by the form of the periodic potential 
if the CdL tunneling occurs. 
\end{abstract}

%%%%%%%%%%%%%%%%%%%%%%

\section{Introduction}
\label{sec:Intro}

In a seminal paper \cite{C&D}, Coleman and de Luccia
studied the vacuum decay 
in the presence of gravity, 
generalizing the earlier work on the false
vacuum decay without gravity \cite{Coleman:1977py}. 
This problem becomes important when considering
the fate of metastable vacua in the early universe 
and the cosmic landscape as suggested by superstring theory \cite{Susskind:2003kw}.
We would like to understand the false vacuum decay 
via nucleation of the bubble of true vacuum within false vacuum,
triggered by a quantum tunneling process.
In the semi-classical analysis, such tunneling process is described by an Euclidean
solution 
called the ``Coleman-de Luccia (CdL) instanton''.

In general, it is very difficult to 
calculate the rate of vacuum decay,   
since the equations describing CdL instantons are highly non-linear. 
Therefore, the CdL tunneling rate has been analyzed 
by using some approximations in most of the previous studies. 
For instance, 
the thin-wall approximation and 
the piecewise-linear potential (``triangular potential'') approximation 
are often used in the literature \cite{TW&Z,Lee:2008hz,Lee:2009bp}.
However, to analyze the false vacuum decay quantitatively,
it is desirable to 
calculate the CdL tunneling rate without using such approximations 
and see whether and when those approximations become good. 

In this paper, we will study the CdL tunneling rate by
directly solving the Euclidean equation of motion
of a single scalar field coupled to gravity, without using
the thin-wall approximation. We consider various types of potential
$V(\phi)$ for the scalar field $\phi$ and evaluate
the tunneling rate numerically.

This paper is organized as follows. 
In section \ref{Sec:CDL} and \ref{Sec:tw}, 
we explain the CdL tunneling and its thin-wall approximation. 
In section \ref{Sec:analysis}, we  numerically calculate the CdL tunneling 
rate for simple potential that have a true and false vacuum.
As examples, we consider $1/\cosh(\phi)$- and $\cos(\phi)$-type potential.
Section \ref{Sec:C&D} is devoted to conclusions and discussions.

%%%%%%%%%%%%%%%%%%%%%%%%

\section{Coleman-de Luccia Tunneling}
\label{Sec:CDL}
\begin{figure}[ht]
  \centering
  \includegraphics[width=7cm,clip]{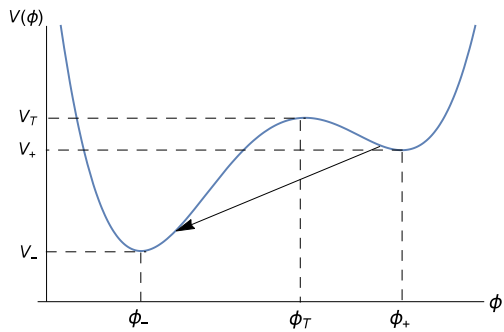}
  \caption{A typical form of potential $V(\phi)$. 
$V_+$ is the false vacuum, $V_-$ is the true vacuum and $V_T$ is the top of the potential. 
The arrow shows tunneling from the false vacuum to true vacuum.}
  \label{fig:potential}
\end{figure}

In this paper, 
we will consider a single scalar field $\phi$ with a potential $V(\phi)$ 
in the presence of gravity, described by the following action:
\begin{equation}
  S = \int d^4x \sqrt{-g} \left[ \frac{1}{2}g^{\mu\nu}\partial_{\mu}\phi\partial_{\nu}\phi 
- V(\phi) - \frac{R}{16\pi G_N} \right] \label{action}
\end{equation}	
where $G_N$ denotes the Newton constant, which is related
to the (reduced) Planck mass $M_{pl}$ by
\begin{equation}
 M_{pl}^2=\frac{1}{8\pi G_N}.
\end{equation}
Let us assume that the potential $V(\phi)$ has two local minima
at $\phi=\phi_{+}$ and $\phi=\phi_{-}$
with potential values $V_{\pm}=V(\phi_{\pm})$.
When $V_{+}>V_{-}$,
the local minima at $\phi=\phi_{+}$ and $\phi=\phi_{-}$
correspond to the false vacuum and true vacuum, respectively.
We also assume that
there is a potential barrier between those two local minima
and the potential has a local maximum
at $\phi=\phi_{T}$ with the potential value
$V_{T}=V(\phi_{T})$ (see Fig. \ref{fig:potential}).
In this paper, we will focus on the transition between two de Sitter vacua
\footnote{In this paper, we do not consider the dS to AdS transitions
in the early universe, since AdS vacuum will eventually end up with a big crunch
and it will never grow to a large universe.} ,
where the potential values of the false vacuum $V_{+}$ and
the true vacuum $V_{-}$
are both positive
\begin{align}
 V_{+}>V_{-}>0.
\end{align}
We are interested in the false vacuum decay
initiated by the nucleation of a bubble of true vacuum
within the false vacuum.
This bubble nucleation occurs via quantum mechanical tunneling process,
which can be described by an
Euclidean bounce solution as discussed in \cite{C&D}.
We assume that the solution is $O(4)$-symmetric.
Taking the ansatz of the metric
\begin{equation}
  ds^2 = d\tau^2 + a(\tau)^2 d\Omega_3^2 \label{metric}
\end{equation}
with  $d\Omega_3^2$ being the metric of a unit 3-sphere,
the  equation of motion reads
\begin{align}
  &\ddot{\phi} + \frac{3\dot{a}}{a}\dot{\phi} = \frac{dV}{d\phi}~, \label{eom1} \\
  &\frac{\dot{a}^2}{a^2} = \frac{1}{a^2} + \frac{1}{3M_{pl}^2}
\left( \frac{1}{2}\dot{\phi}^2 - V  \right) \label{eom2}~.
\end{align}
Here we also assumed that $\phi$ depends only on $\tau$,
and the dots represent the derivative with respect to $\tau$.
Using \eqref{eom1} and \eqref{eom2}, we find
\begin{equation}
  \ddot{a} = -\frac{a}{3M_{pl}^2}\left( \dot{\phi}^2 + V \right) \label{eom3}~.
\end{equation}
We can choose \eqref{eom1} and \eqref{eom3}
as the independent set of equations, and we would like to solve this set of equations.
The $O(4)$-symmetric bounce solution is topologically a 4-sphere,
with vanishing size of $S^3$ at $\tau=0$ and $\tau=\tau_\text{max}$:
\begin{align}
 a(0)=a(\tau_\text{max})=0.
\end{align}
To avoid the conical singularity at $\tau=0$
and $\tau=\tau_\text{max}$,
we require that $a(\tau)$ behaves as
\begin{align}
& \lim_{\tau\to0}a(\tau)=\tau+\mathcal{O}(\tau^2),\nonumber\\
& \lim_{\tau\to\tau_\text{max}}a(\tau)=(\tau_\text{max}-\tau)+\mathcal{O}((\tau_\text{max}-\tau)^2).
\label{a-reg}
\end{align}

In semi-classical (small $\hbar$) approximation, the
tunneling rate per unit volume is written as
\begin{align}
  \Gamma &= Ae^{-B}. \label{DR}
\end{align}
We will refer to $B$ as the ``bounce factor'',
which is determined by the on-shell action of the Euclidean
the bounce solution, while the
coefficient $A$ comes from the one-loop determinant around this solution. 
We assume that  $A$ is of order  $\mathcal{O}(1)$,
and we will focus on the computation of bounce factor $B$ in what follows.
As shown in \cite{C&D},
the bounce factor $B$ is given by
\begin{align}
 B &= S_E(\phi) - S_E(\phi_+)~\label{b},
\end{align}
where $S_E(\phi)$ is the on-shell action for the bounce solution
 of \eqref{eom1} and \eqref{eom3}
\begin{equation}
  S_E = 4\pi^2\int_0^{\tau_{\text{max}}} d\tau \left[ a(\tau)^2V(\phi(\tau)) 
- 3M_{pl}^2a(\tau) \right], \label{Eaction}
\end{equation}
while
$S_E(\phi_+)$ is the Euclidean action of the reference solution,
in which the scalar field is sitting
at the false vacuum $\phi=\phi_+$.
In this reference solution, the equation \eqref{eom3}
for
$a(\tau)$ reduces to a simple harmonic oscillator equation
\begin{align}
 \ddot{a}=-\omega^2 a,\qquad  \omega^2 = \frac{V_{+}}{3M_{pl}^2}, 
\end{align}
and the reference solution of false vacuum is easily obtained as
\begin{align}
\begin{aligned}
  \phi(\tau) = \phi_+,\qquad
  a(\tau) = \frac{1}{\omega}\sin{\omega\tau}~.  \\
\end{aligned}
\label{sol.o.fv}
\end{align}
One can  see that this reference solution obeys the 
regularity condition \eqref{a-reg}
at $\tau=0$ and $\tau=\tau_{\text{max}}$
with
\begin{align}
 \tau_{\text{max}} = \frac{\pi}{\omega}. 
\end{align}
The action of the reference solution at false vacuum is
 evaluated as
\begin{equation}
  S_E(\phi_+) = -\frac{24\pi^2M_{pl}^4}{V_{+}} 
\label{ac.o.fv}~.
\end{equation}

For general solution, eq.\eqref{eom3}
can be seen as a harmonic oscillator
with field dependent frequency.
On the other hand, eq.\eqref{eom1}
for the scalar field has a friction term
proportional to $\dot{\phi}$, hence
there is no simple way to find an integral of motion. 
Except for very special potentials \cite{D&H,Kanno:2011vm}, 
we do not have analytic solutions of \eqref{eom1}
and \eqref{eom3} for generic choice of potential.
Instead, we will analyze the system of equations \eqref{eom1}
and \eqref{eom3} numerically.

However, it turns out that this is not so straightforward as one might think. 
We are looking for a solution
approaching 
\begin{align}
 \lim_{\tau\to0}\phi(\tau)=\phi_\text{ini},\qquad
\lim_{\tau\to\tau_\text{max}}\phi(\tau)=\phi_\text{fin}
\end{align}
and obeying the regularity condition for the metric \eqref{a-reg}.
One difficulty is that the value of $\tau_\text{max}$
at which the factor $a(\tau)$ vanishes is not known a priori, and 
$\tau_\text{max}$ is determined by the solution itself.
In other words, we can only know the value of $\tau_\text{max}$ after we solved the  equations of motion \eqref{eom1} and
\eqref{eom3}, and the boundary condition at $\tau_\text{max}$ cannot be simply set
from the beginning.
This makes the construction of a regular solution a highly 
non-trivial problem.
Another difficulty is that, in general
the initial and the final values of the
scalar field are different from the
local minima of the potential
\begin{align}
 \phi_\text{ini}\not=\phi_{+},\qquad
\phi_\text{fin}\not=\phi_{-}.
\label{diff-phi}
\end{align}
The existence of a regular solution is not guaranteed
if we choose the initial value $\phi_\text{ini}$ at random.
As in the case of $\tau_\text{max}$,
we do not know the initial value $\phi_\text{ini}$
of scalar field a priori.
To find a regular solution, 
we have to fine-tune the initial value 
$\phi_\text{ini}$ so that the solution obeying the conditions
\eqref{a-reg}  and \eqref{diff-phi} exists.

The fact \eqref{diff-phi} that the tunneling
does not necessarily start from the minimum of the potential
has a natural
interpretation as the effect of the thermal fluctuation
due to the Gibbons-Hawking temperature associated with the de Sitter horizon \cite{H&W}.
A physical picture is that the scalar field first climbs up 
the potential from $\phi_{+}$ to $\phi_\text{ini}$
by thermal fluctuation, then it tunnels through the barrier.

In \cite{H&W,Banks:2002nm,Lavrelashvili:2006cv,Lee:2011ms,Lee:2014ula},
it is found that there are oscillating solutions 
crossing the top of the potential barrier $k$ times $(k=1,2,\cdots)$
before settling down to $\phi_\text{fin}$.
It is argued that as the number of oscillation $k$ increases
$B$ becomes larger, which implies that the dominant contribution
comes from the non-oscillating $(k=1)$ solution. 
Thus, in this paper we will consider
the non-oscillating solution only.

The numerical study of bounce solution can be found in the literature
\cite{Lee:2013mza,Samuel:1991dy,Samuel:1991mz,ZSY&S,Kachru:2003aw,Casadio:2011jt,
Isidori:2007vm,Artymowski:2015mva,Espinosa:2015zoa}, 
but the systematic study for various types of
potential have not been performed, as far as we know.
In this paper, we will consider two types of potentials, as simple examples.
We should stress that the our numerical construction of solutions
can be applied to arbitrary shape of the potential.

We find that as the initial value $\phi_\text{ini}$ approaches
the local maximum $\phi_T$,
the bounce factor $B$ of our solution
is well approximated by the bounce factor $B_\text{HM}$ of the Hawking-Moss (HM) tunneling \cite{HM},
\begin{equation}
  B_{HM} = S_E(\phi_T) - S_E(\phi_+) 
= 24\pi^2M_{pl}^4\left( -\frac{1}{V_T} + \frac{1}{V_+} \right)~, \label{bhm}
\end{equation}
corresponding to the solution sitting at the top of the barrier $\phi=\phi_T$.

%%%%%%%%%%%%%%%%%%%%%%

\section{The thin-wall approximation}
\label{Sec:tw}

The thin-wall approximation requires a rapid transition of $\phi$ from $\phi_-$ to $\phi_+$
near the wall. 
It is argued in \cite{C&D}
that the thin-wall approximation is valid
when the potential difference $V_{+}-V_{-}$
between the true vacuum and the false vacuum is small. 

To compute the bounce factor $B_\text{tw}$
in the thin-wall approximation,
we divide the integration region into three parts:
 outside and inside of the bubble, and the contribution of the wall.
In the thin-wall approximation, we can approximate
that  $\phi$ is sitting at the false vacuum $\phi=\phi_+$ outside the bubble.
Thus the contribution outside the bubble is
\begin{equation}
  B_{\text{out}} = S_E(\phi_+) - S_E(\phi_+) = 0. \label{Bout}
\end{equation}
Near the wall, we can neglect the friction term in  \eqref{eom1}
\begin{equation}
  \frac{\dot{a}}{a} \dot{\phi} \ll 1, 
\end{equation}
and \eqref{eom1} becomes
\begin{equation}
  \ddot{\phi} \simeq \frac{dV}{d\phi}~.
\end{equation}
This equation can be solved as:
\begin{equation}
  \dot{\phi} \simeq \sqrt{2[V(\phi)-V(\phi_{\pm})]}. \label{dotphi}
\end{equation}
Therefore, the contribution of the wall, $B_{\text{wall}}$, is
given by
\begin{align}
  B_{\text{wall}} &= 2\pi^2a^3 \left( \int_{\phi_T}^{\phi_+} d\phi \sqrt{2(V(\phi)-V_+)} + \int_{\phi_-}^{\phi_T} d\phi \sqrt{2(V(\phi)-V_-)} \right) \notag\\
                     &\equiv 2\pi^2a^3T. \label{Bwall}
\end{align}
where $a$ is the bubble size 
and $T$ is the tension of the wall which is determined by the barrier 
between the false vacuum and the true vacuum.
Inside the bubble, $\phi$ is sitting approximately at the true vacuum. 
For a constant $\phi$, eq.\eqref{eom2} becomes 
\begin{equation}
  d\tau = da \left( 1 - \frac{a^2V(\phi)}{3M_{pl}^2} \right)^{-1/2} \notag~.
\end{equation}
Therefore, eq.\eqref{Eaction} is written as
\begin{equation}
  S_{E,\text{in}}(\phi) =  \frac{12\pi^2M_{pl}^4}{V(\phi)} \left( \left( 1 - \frac{aV(\phi)}{3M_{pl}^2} \right)^{3/2} - 1 \right)~.
\end{equation}
Hence, the contribution of the inside of the bubble is
\begin{align}
  B_{\text{in}} &= S_E(\phi_-) - S_E(\phi_+) \notag\\
                  &= 12\pi^2M_{pl}^4 \left[ \frac{1}{V_-}\left( \left( 1 - \frac{aV_-}{3M_{pl}^2} \right)^{3/2} - 1 \right) - \frac{1}{V_+}\left( \left( 1 - \frac{aV_+}{3M_{pl}^2} \right)^{3/2} - 1 \right) \right] \label{Bin}~. 
\end{align}
Adding the three contributions $B_\text{out}$,
$B_\text{wall}$, and $B_\text{in}$,
we obtain
\begin{align}
  B &= 2\pi^2a^3T \notag\\
     &~~~~~+ 12\pi^2M_{pl}^4 \left[ \frac{1}{V_-}\left( \left( 1 - \frac{a^2V_-}{3M_{pl}^2} \right)^{3/2} - 1 \right) - \frac{1}{V_+}\left( \left( 1 - \frac{a^2V_+}{3M_{pl}^2} \right)^{3/2} - 1 \right) \right]~\label{allB}.
\end{align}
The bubble size $a_0$ is determined by extremizing  $B$ in  \eqref{allB}
\begin{equation}
  \frac{\partial B}{\partial a} \bigg{|}_{a=a_0} = 0 \label{condition}~,
\end{equation}
from which $a_0$ is found to be
\begin{equation}
  \frac{1}{a_0^2} = \frac{(V_+ - V_-)^2}{9T^2} + \frac{V_++V_-}{6M_{pl}^2} + \frac{T^2}{16M_{pl}^4}. \label{a0}
\end{equation}
Plugging $a_0$ in \eqref{a0} into \eqref{allB},
the bound factor $B_{\text{tw}}$ in the thin-wall approximation is
given by
\begin{align}
  B_{\text{tw}} &= 2\pi^2a_0^3T \notag\\
                    &~~~~+ 4\pi^2M_{pl}^2 \left[ \frac{V_-}{3M_{pl}^2}\left( \left( 1 - \frac{3M_{pl}^2a_0^2}{V_-} \right)^{3/2} - 1 \right) - \frac{V_+}{3M_{pl}^2}\left( \left( 1 - \frac{3M_{pl}^2a_0^2}{V_+} \right)^{3/2} - 1 \right) \right] \label{btw}.
\end{align}

%%%%%%%%%%%%%%%%%%%%

\section{Numerical Analysis}
\label{Sec:analysis}

In this section, we construct the CdL instanton solution numerically for
some scalar potentials,
by solving the set of equations \eqref{eom1} and \eqref{eom3}.
Then we compute the bounce factor $B$ from this numerical solution and
study the dependence of $B$ on the various parameters in the potential.
As simple examples,
we consider the $1/\cosh(\phi)$-type potential and the $\cos(\phi)$-type periodic potential.
They have natural cosmological and phenomenological applications.  
We should stress that our numerical approach can be applied to
any shape of potential, say the $\phi^4$ potential with a negative mass-squared term
(double-well potential).

To solve \eqref{eom1} and \eqref{eom3},
we set the initial conditions of $a(\tau)$ and $\phi(\tau)$ as
\begin{equation}
  a(0) = 0~,~\dot{a}(0) = 1~,~\phi(0) = \phi_{\text{ini}}~,~\dot{\phi}(0) = 0. \label{inicond}
\end{equation}
There is one subtle point: To find a solution numerically, 
we have to introduce a regularization parameter
for the initial value of the scale factor $a(\tau)$
\begin{align}
 a(0) =\varepsilon,\qquad 0<\varepsilon\ll1,
\end{align}
so that the denominator $\frac{\dot{a}}{a}\dot{\phi}$ in the friction term
of \eqref{eom1} is non-zero at the initial value.
We choose the value of the parameter
$\varepsilon$ as
\begin{align}
 \varepsilon=10^{-100}.
\end{align}
Once we know the numerical bounce solution,
we can numerically calculate the bounce factor $B$
by plugging the solution into \eqref{b}. 
Then, we can compare this result with the thin-wall approximation
$B_{\text{tw}}$ \eqref{btw} and the bounce factor of Hawking-Moss instanton $B_{\text{HM}}$ \eqref{bhm}.

\subsection{$1/\cosh$ potential} 

First, we consider the potential $V(\phi)$ of the form
\begin{equation}
  V(\phi) = v_0 + \frac{v_1}{\cosh{(\phi/m)}} + v_2\tanh{(\phi/m)}, \label{coshtanh}
\end{equation}
where $m$ is a mass parameter, and 
$v_0$, $v_1$ and $v_2$ \eqref{coshtanh} are arbitrary parameters.
This type of $1/\cosh$ potential naturally appears 
from the open string tachyon living on unstable D-branes
and it has an application to tachyon driven cosmology
\cite{Sen:2002qa,Sen:2003mv}. We also
added a term $\tanh\phi$ to make the true vacuum and false vacuum to have
slightly different potential values.
We will solve the set of equations \eqref{eom1} and
\eqref{eom3} with this potential. 
In doing this, we set $m/M_{pl}=1/30$ for definiteness.

\begin{figure}[htbp]
  \centering
  \includegraphics[width=9cm,clip]{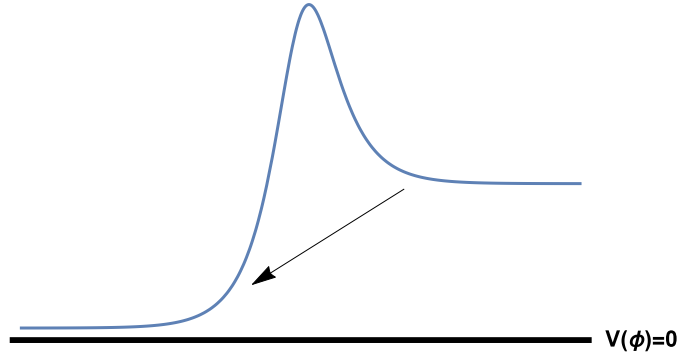}
  \caption{An example of the potential \eqref{coshtanh}. 
The arrow shows tunneling from the false vacuum to the true vacuum.}
  \label{fig:coshtanh}
\end{figure}

The shape of the potential \eqref{coshtanh} is shown in Figure \ref{fig:coshtanh}.
We analyze the CdL tunneling rate by changing the parameters $v_0$, $v_1$ and $v_2$, 
which is equivalent to changing the values $V_T$, $V_+$ and $V_-$. 
To have the transition between two de Sitter vacua, 
we take the potential to be always positive: $V(\phi)>0$.

\begin{figure}[!h]
\begin{center}
  \begin{tabular}{cc}
    \includegraphics[width=8cm,clip]{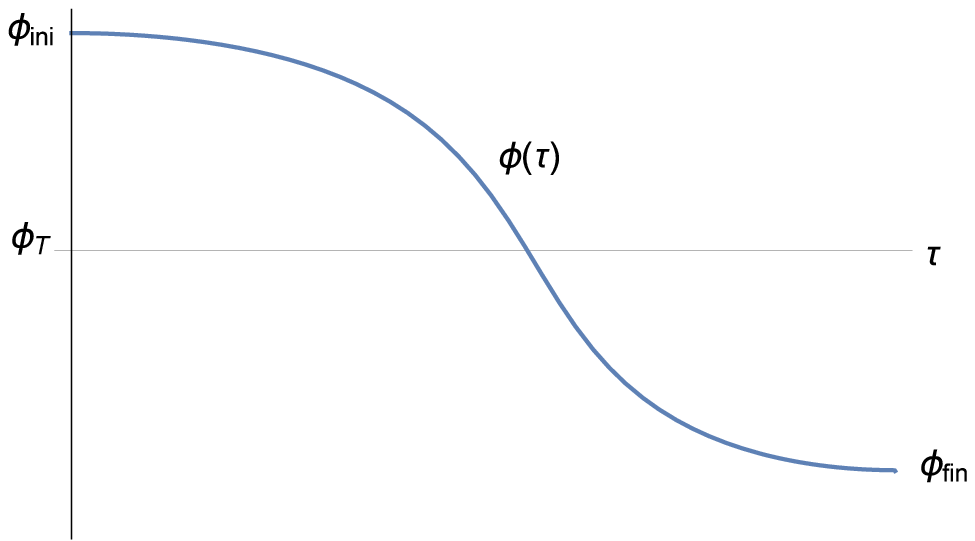} &
    \includegraphics[width=8cm,clip]{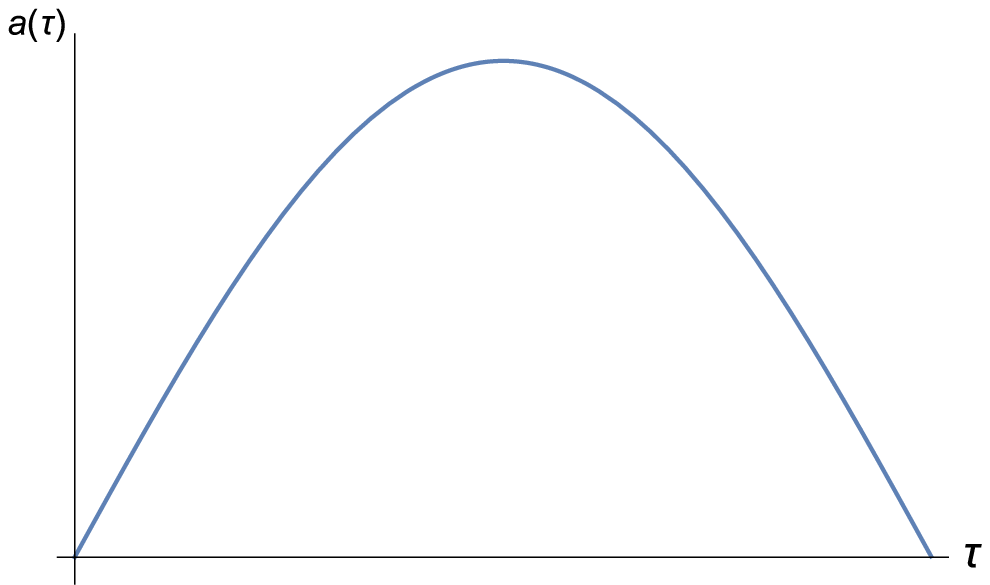}
  \end{tabular}
 \caption{Left: A solution of scalar field for CdL instanton
from $\tau=0$ to $\tau=\tau_{\text{max}}$. 
$\phi(\tau)$ is $\phi_{\text{ini}}$ at $\tau=0$ 
and $\phi_{\text{fin}}$ at $\tau=\tau_{\text{max}}$. 
Right: The associated solution of scale factor $a(\tau)$ 
from $\tau=0$ to $\tau=\tau_{\text{max}}$.}
 \label{fig:phi&a}
\end{center}
\end{figure}

In this case, the false vacuum and the true vacuum correspond
 to $\phi_{\pm}=\pm\infty$.
 However, as discussed around \eqref{diff-phi}
the scalar field
 in the bounce solution starts and ends at finite values.  
An example of the shape of the solution  
$\phi(\tau)$ and $a(\tau)$ are shown in Figure \ref{fig:phi&a}. 
From this figure, one can see that the actual solution of CdL instanton does not 
start and end at $\phi=\pm\infty$, and the initial and the final values of
the scalar field are some finite values.
The scale factor $a(\tau)$ is not much different from the reference solution
in \eqref{sol.o.fv}, in this example.

We calculate $B$ and $B_{\text{HM}}$ for various values of $V_T$ with fixed 
potential difference $V_+-V_-$ and $V_T-V_{+}$.
In other words, we change the off-set of the potential $v_0$
in \eqref{coshtanh}
without changing the shape of the potential parametrized by $v_1,v_2$.
\begin{figure}[!h]
\begin{center}
  \begin{tabular}{cc}
    \includegraphics[width=8cm,clip]{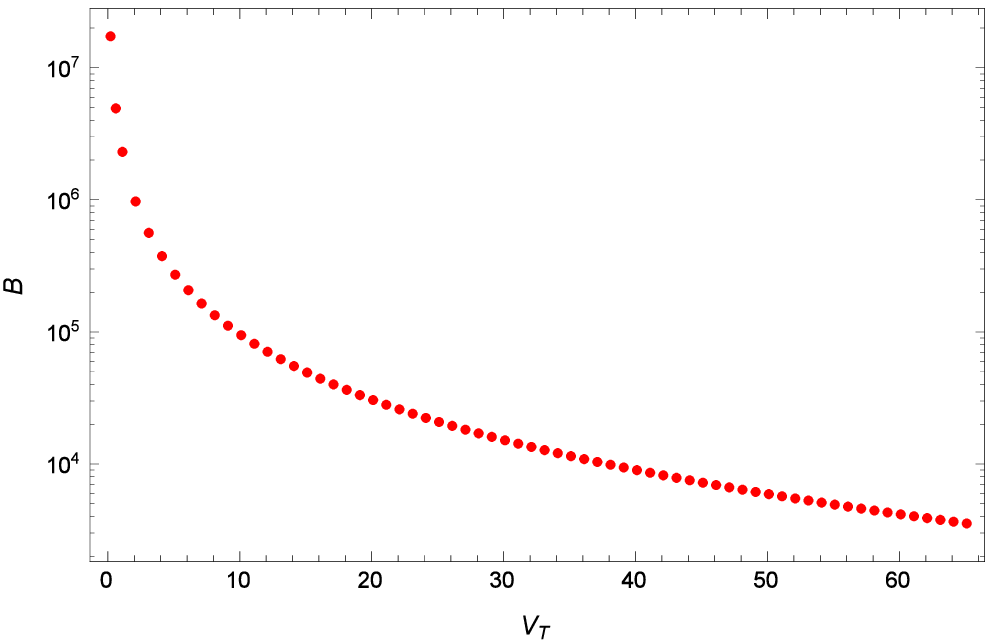} &
    \includegraphics[width=8cm,clip]{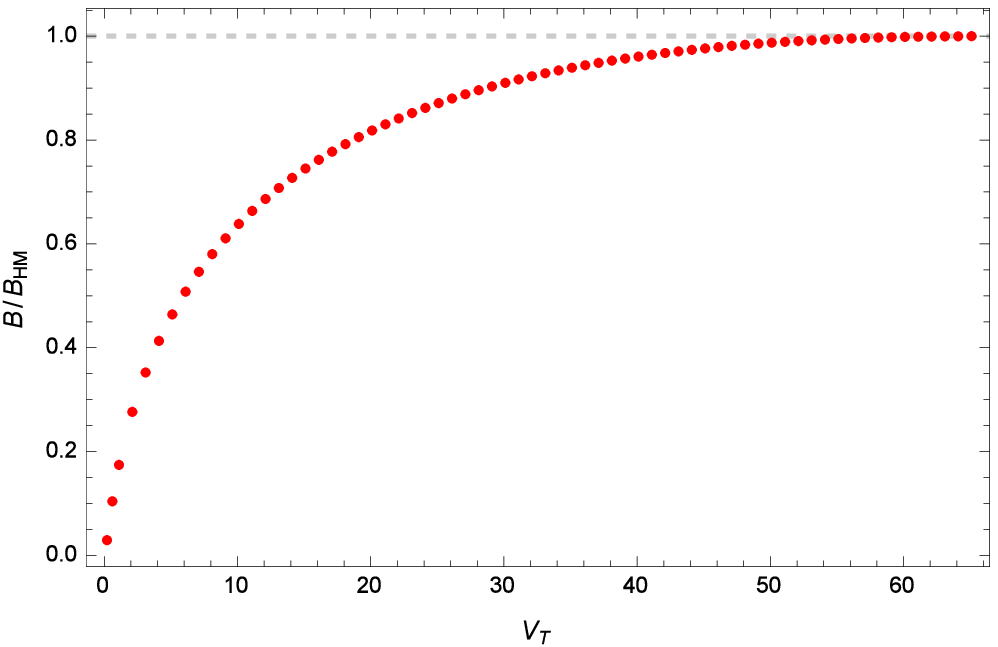}
  \end{tabular}
 \caption{Left is the plot of $B$ against $V_T$. 
Right is the plot of $B/B_{\text{HM}}$ against $V_T$. 
In the potential \eqref{coshtanh},
we change the value of $v_0$ 
with the parameters $v_1,v_2$ fixed to $v_1=1/10$ and $v_2=1/40$ , 
In other words, we change $V_T$ with fixed $V_T-V_+$ and $V_+-V_-$.}
 \label{fig:coshtanhvT}
\end{center}
\end{figure}
In Figure \ref{fig:coshtanhvT}, we show the plot of $B$ and $B/B_\text{HM}$ 
as a function of $V_T$.
From Figure \ref{fig:coshtanhvT}, one can see that 
$B$ decreases as $V_T$ increases, and $B$ approaches $B_\text{HM}$ for large $V_T$. 
This shows that the initial value $\phi_\text{ini}$ of
scalar field approaches  
the top of the potential barrier when $V_T$ becomes large. 
On the other hand, $\phi_{\text{ini}}$ approaches $\phi_+$ 
when $V_T$ is small.
Therefore, the probability of the CdL tunneling is controlled by
the off-set scale of the potential in this setup.  

This suggests the following picture: When the potential height becomes large the 
tunneling probability increases,
and the universe is very mobile and tends to
move around the landscape of vacua.
On the other hand, as the energy scale becomes small,
the tunneling to other vacua is highly suppressed
and the universe is bound to one of the vacua.
This is analogous to the metal-insulator transition in
 the condensed-matter physics.
In the next subsection, we will consider the tunneling rate
in the periodic potential.
The analogy to the electron in the crystal 
is more appropriate in the example of periodic potential,
as advocated in \cite{TW&Z}.

%%%%%%%%%%%%%%%%%%%

\subsection{Periodic Potential}
\label{Sec:periodic}

Next, we analyze the tunneling rate in the case of periodic potential
\begin{equation}
  V(\phi) = \Lambda^4 \left( v_0 + v_1\cos(\phi/f) \right), \label{cos}
\end{equation}
where $\Lambda$ is some non-perturbatively generated scale, 
$f$ is the axion decay constant, $v_0$ and $v_1$ are arbitrary parameters.
This type of potential was first introduced to solve
the strong CP problem by promoting
the theta-angle of QCD to a dynamical field, called axion \cite{Peccei:1977hh,Peccei:1977ur}.
This type of axionic scalar fields is considered in many contexts, 
beyond the original QCD axion.
In particular, axionic fields appear naturally 
in string compactifications from the extra dimensional components
of the anti-symmetric tensor fields in string theory. 

\begin{figure}[htbp]
  \centering
  \includegraphics[width=9cm,clip]{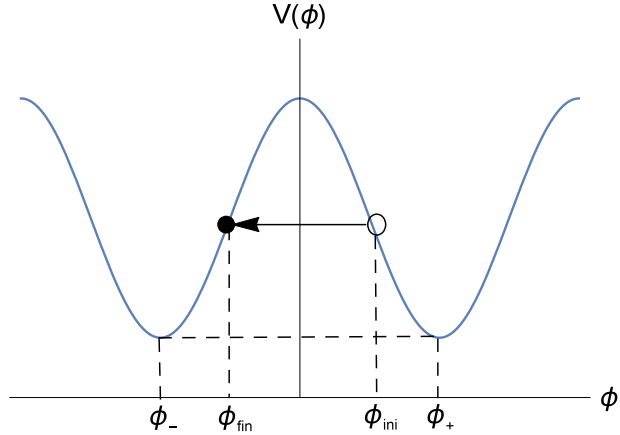}
  \caption{An example of the potential \eqref{cos}. 
The arrow shows tunneling from $V(\phi_{\text{ini}})$ to $V(\phi_{\text{fin}})$.}
  \label{fig:axion_potential}
\end{figure}

The shape of the potential \eqref{cos} is shown in Figure \ref{fig:axion_potential}.
\begin{figure}[!h]
\begin{center}
  \begin{tabular}{cc}
    \includegraphics[width=8cm,clip]{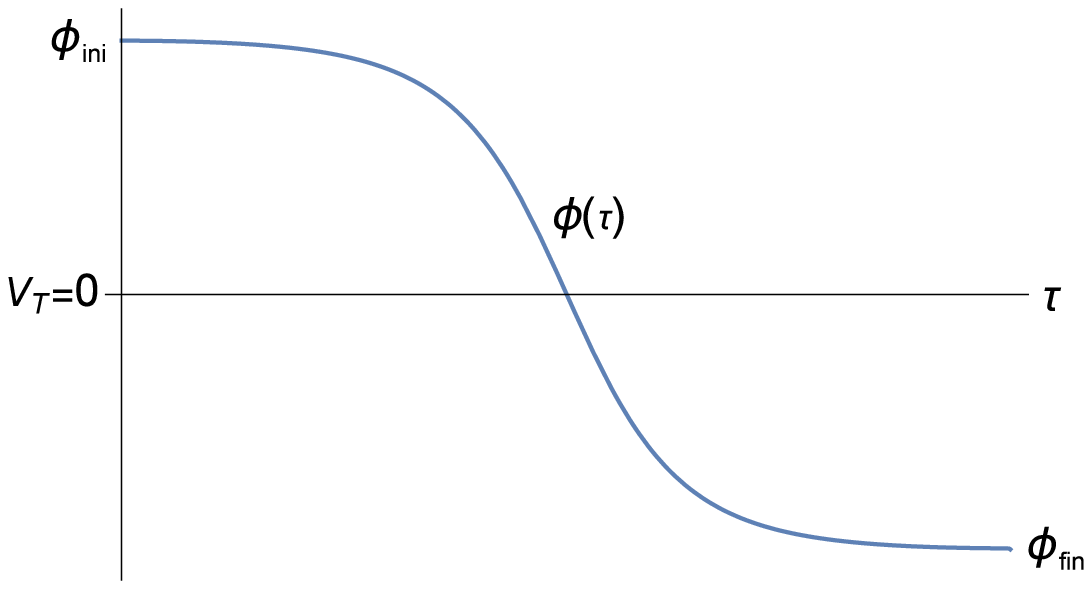} &
    \includegraphics[width=8cm,clip]{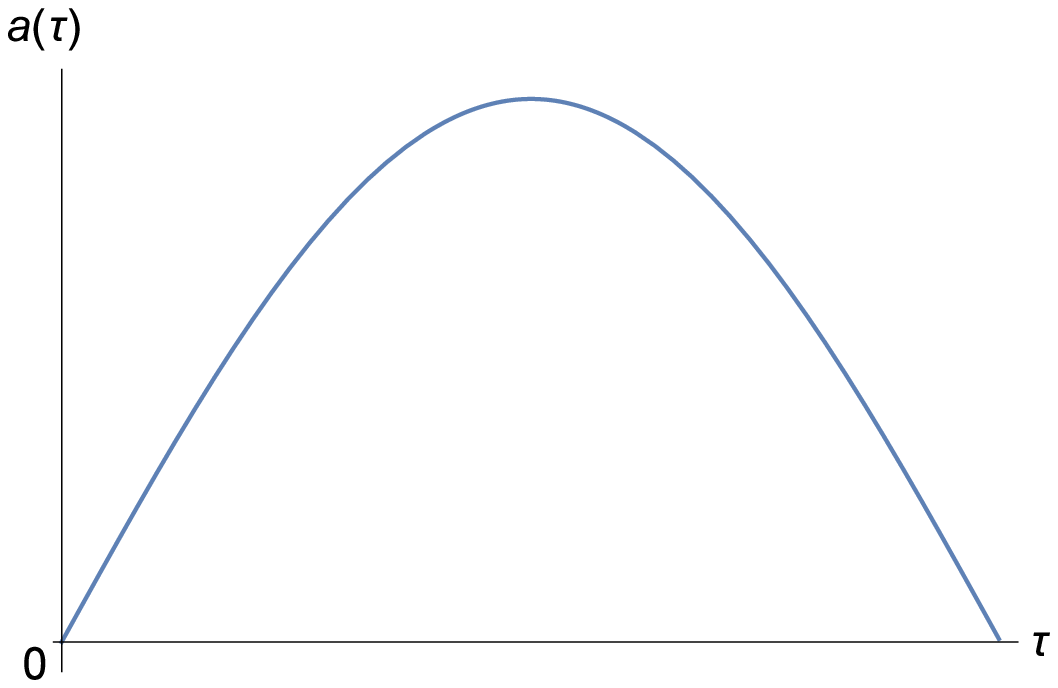}
  \end{tabular}
 \caption{Left: A solution of scalar field for the CdL instanton
from $\tau=0$ to $\tau=\tau_{\text{max}}$.
$\phi(\tau)$ is $\phi_\text{ini}$ at $\tau=0$ and $\phi_\text{fin}$ at $\tau=\tau_{\text{max}}$. 
Right: The associated solution of scale factor $a(\tau)$ 
from $\tau=0$ to $\tau=\tau_{\text{max}}$.}
 \label{fig:axion_phi&a}
\end{center}
\end{figure}
The procedure of solving the equations of motion \eqref{eom1} and
\eqref{eom2} are  the same as the previous subsection. 
Using the numerical solution of $a(\tau)$ and $\phi(\tau)$, we evaluate
the bounce factor $B$ for various values of the parameters in \eqref{cos}.

\begin{figure}[!h]
\begin{center}
  \begin{tabular}{cc}
    \includegraphics[width=8cm,clip]{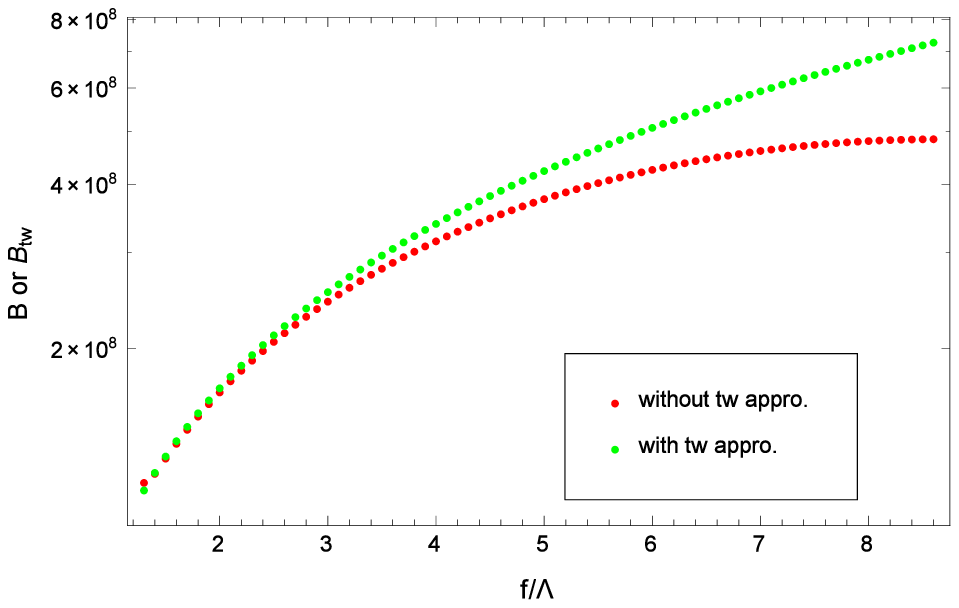} &
    \includegraphics[width=8cm,clip]{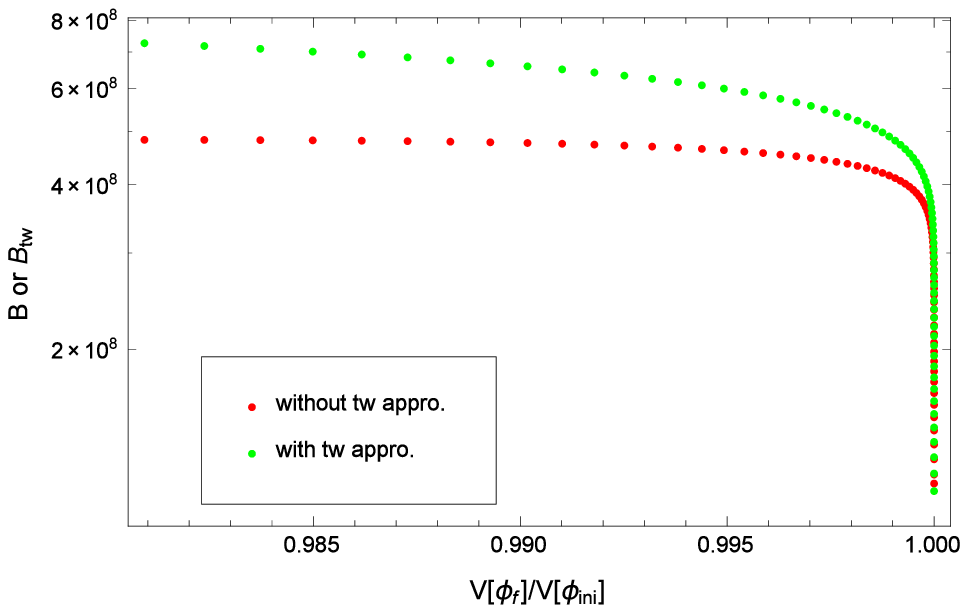}
  \end{tabular}
 \caption{Left is the plot of $B$ or $B_{\text{tw}}$ against  $f/\Lambda$. 
Right is the plot of $B$ or $B_{\text{tw}}$ against $V_+/V(\phi_{\text{ini}})$.
Red dots are the result of eq.\eqref{b} and green dots are its of eq.\eqref{btw} 
in both figure. 
In this figure we have fixed $V_T=1$, $V_T-V_+=1/200$ and $\Lambda/M_{pl}=1/100$.}
 \label{fig:VT1_Mp100}
\end{center}
\end{figure}
First, by fixing $V_T = 1.0$ and $V_T-V_+ = 1/200$, $\Lambda/M_{pl}=1/100$,   
we calculate $B$ and $B_{\text{tw}}$ for  various values of the axion decay constant $f$. 
Figure \ref{fig:VT1_Mp100} shows the plot of $B$ (or $B_{\text{tw}}$) 
against  $f$ (left figure) 
and $V_+/V(\phi_{\text{ini}})$ (right figure).
From the left figure of Fig. \ref{fig:VT1_Mp100}, 
one can see that both $B$ and $B_{\text{tw}}$ become large as $f$ increases. 
Also, this figure shows that the thin-wall approximation gets worse when 
$f$ becomes large.
We also find that when $f$ increases, the initial value of the potential
$V(\phi_{\text{ini}})$ approaches the 
value of the potential $V_T$ at the top of the barrier. 
This implies that the bounce factor $B$ approaches the Hawking-Moss value $B_{\text{HM}}$
for large $f$.
Interestingly, in order for the regular solution to exist,
the range of allowed $f$ is restricted
\begin{align}
 f_\text{min}<f<f_\text{max},
\label{range-f}
\end{align} 
since the initial value of the potential is bounded from both above and below:
\begin{align}
 V_+<V(\phi_{\text{ini}})<V_T.
\end{align}
The bounds $f_\text{min}$ and $f_\text{max}$ in \eqref{range-f}
correspond to the limit where $V(\phi_{\text{ini}})$
approaches $V_{+}$ and $V_T$, respectively:
\begin{align}
 \lim_{f\to f_\text{min}}V(\phi_{\text{ini}})=V_+,\qquad
\lim_{f\to f_\text{max}}V(\phi_{\text{ini}})=V_T.
\end{align}
For the parameter choice in Figure \ref{fig:VT1_Mp100}, we find 
that the allowed range of $f$ is $1.3 \leq f/\Lambda \leq 8.6$.
It would be interesting to understand the cosmological implication (if any)
of this bound on the axion decay constant. 

\begin{figure}[!h]
\begin{center}
  \includegraphics[width=8cm,clip]{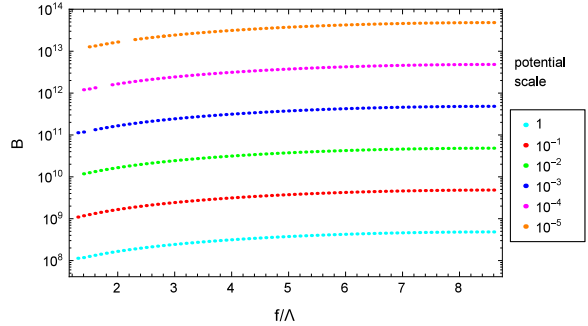}
  \caption{This figure is the plot of $B$ against $f/\Lambda$ for each of the potential scale. 
The light blue, red, green, blue, pink and orange dots correspond to 
$V_T=1$, $10^{-1}$, $10^{-2}$, $10^{-3}$, $10^{-4}$ and $10^{-5}$, respectively. 
In this figure we have fixed $(V_T-V_{\pm})/V_T=1/200$ and $\Lambda/M_{pl}=1/100$.}
 \label{fig:Mp100_VT}
\end{center}
\end{figure}

Next, by fixing $(V_T-V_+)/V_T=1/200$,    
we calculate the bounce factor $B$ by varying the axion
decay constant $f$ for various values of  $V_T$. 
\begin{figure}[!h]
\begin{center}
  \includegraphics[width=8cm,clip]{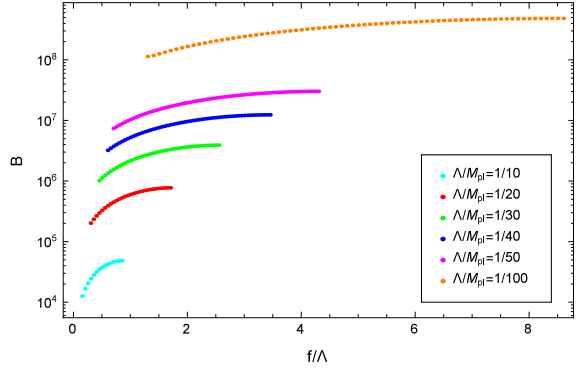}
  \caption{This figure is the plot of $B$ against $f/\Lambda$. 
The light blue, red, green, blue, pink and orange dots correspond to 
$\Lambda/M_{pl}=1/10$, $1/20$, $1/30$, $1/40$, $1/50$ and $1/100$, respectively. 
In this plot, we have fixed $V_T=1$ and $(V_T-V_{\pm})=1/200$.}
 \label{fig:VT1_Mp}
\end{center}
\end{figure}
Figure  \ref{fig:VT1_Mp} and Figure \ref{fig:Mp100_VT}
are the plot $B$ against $f/\Lambda$
with various values of $\Lambda/M_{pl}$ or $V_T$, respectively.
From those figures,
one can see that $B$ becomes large as the energy scale of the axion 
potential $\Lambda/M_{pl}$ becomes small.
From Figure \ref{fig:Mp100_VT}, we can also read off the behavior of $B$ that
when the potential scale is reduced by a factor of $10^{-1}$,  
the bounce factor $B$ roughly becomes $10$ times larger. 

This behavior suggests the analogy with the electrons in a crystal
as mentioned at the end of the previous subsection.
As the energy scale of the axion 
potential $\Lambda/M_{pl}$ is lowered, 
the tunneling probability becomes negligibly small and the universe
is ``tight-binded'' to one of the vacua.
On the other hand, as the energy scale increases,
the wavefunction of the universe
spreads over the landscape of vacua
as a Bloch-wave in analogy with the conducting electrons in a metal \cite{TW&Z}.
From Figure \ref{fig:VT1_Mp},
we also notice that the allowed range of $f$ \eqref{range-f} changes as we
change the energy scale of axion potential $\Lambda$.
In this example, the maximum of the allowed 
axion decay constant $f_\text{max}$ never exceeds the Planck scale
$M_{pl}$.

%%%%%%%%%%%%%%%%%%%

\section{Conclusion and Discussion}
\label{Sec:C&D}

In the paper, we numerically analyzed
the false vacuum decay in the presence of gravity described by the CdL instanton,
and studied the dependence of the bounce factor $B$ in \eqref{b} on
the various parameters in the scalar potential.
We also numerically calculated the bounce factor $B_{\text{tw}}$  in the thin-wall approximation, 
and compared $B$ with $B_{\text{tw}}$. 
We find that when the initial value of the potential $V(\phi_{\text{ini}})$
becomes large 
the thin-wall approximation gets worse. 
Therefore, the thin-wall approximation can only be trusted in the case 
that $V(\phi_{\text{ini}})$ is close to the false vacuum value $V_{+}$, and
the difference of the potential between the false vacuum and the true vacuum is small 
$V_+-V_-\simeq0$.

As simple examples, we considered the $1/\cosh$ potential
and the axionic periodic potential, where both types of 
potential have an interesting application to cosmology.
Especially, the periodic axionic potentials have been considered in many cosmological models, such 
as natural inflations \cite{Freese:1990rb} 
and the cosmic landscape in superstring compactifications. 
We found that the CdL tunneling rate of this axionic potential 
depends on the energy scale $\Lambda$ 
and the axion decay constant $f$ in \eqref{cos}.
We find that the tunneling probability is highly suppressed
when $\Lambda$
becomes small. 
From the observation we know that
the cosmological constant of our universe is very small, hence
the value of the scalar potential of our current universe is expected to be at very low energy,  
hence the tunneling to other vacuum can be safely ignored in today's universe.
On the other hand, the tunneling effect 
cannot be ignored when we consider the physics around the Planck scale.
We also find that the axion decay constant $f$ is bounded both from
above and below \eqref{range-f},
in order for the smooth tunneling solution to exist. It would be interesting to
see if this bound has some cosmological implications.

%%%%%%%%%%%%%%%%%%

%\section*{Acknowledgments}

%%%%Reference%%%%%

\end{document}